\documentclass[french]{gretsi}
\usepackage[utf8]{inputenc}
\usepackage[T1]{fontenc}
\usepackage[bookmarks,hypertexnames=false,debug,pdfborderstyle={/S/U/W 1}]{hyperref}
\usepackage{amsmath, amssymb, amsfonts}
\usepackage[
	maxbibnames=1,				
	citestyle=numeric-comp,		
    sorting=none,
    backend=biber]				
{biblatex}						

\newtheorem{hypo}{Hypothèse}
\newtheorem{prop}{Proposition}
\usepackage{stmaryrd}   
\RequirePackage{stfloats} 
\RequirePackage{subcaption} 
\RequirePackage{epstopdf}
\usepackage{mathtools}
\usepackage{xcolor} 
\definecolor{ao}{rgb}{0.0, 0.0, 1.0}
\usepackage[version=3]{mhchem}
\DeclareMathOperator*{\ms}{\textrm{ }}
\DeclareMathOperator*{\Espe}{\mathbb{E}}
\DeclareMathOperator*{\CZU}{{1\gamma}^{511}_{\textrm{COR}}}     
\DeclareMathOperator*{\CZD}{{2\gamma}^{}_{\textrm{LOR}}}     
\DeclareMathOperator*{\CUZ}{{1\gamma}^{1157}_{\textrm{COR}}}     
\DeclareMathOperator*{\CUU}{{2\gamma}^{}_{\textrm{COR}}}     
\DeclareMathOperator*{\CUD}{{3\gamma}^{}_{}}     

\def\ie{\textit{i.e.} }
\def\cf{cf. }

\begin{document}
\titre{Reconstruction multiclasse pour l'imagerie TEP 3-photons : application à la caméra XEMIS2}   
\auteurs{
  \auteur{Mehdi}{Latif}{mehdi.latif@ls2n.fr}{1,2}
  \auteur{Jérôme}{Idier}{jerome.idier@ls2n.fr}{1}
  \auteur{Thomas}{Carlier}{thomas.carlier@chu-nantes.fr}{2}
  \auteur{Simon}{Stute}{simon.stute@chu-nantes.fr}{2}
}
\affils{
  \affil{1}{Nantes Université, École Centrale Nantes, LS2N, CNRS, UMR 6004, F-44000 Nantes, France}
  \affil{2}{Nantes Université, CHU Nantes, CRCI2NA, F-44000 Nantes, France}
}
\resume{ 
Dans cette contribution, nous abordons le problème de reconstruction d’image de distribution radioactive pour lequel l’information disponible provient de plusieurs classes de données distinctes, chacune associée à une combinaison spécifique de détections . Nous présentons un cadre théorique permettant de mesurer l'apport informationnel de chaque classe et nous développons un algorithme itératif dédié à la reconstruction multiclasse basé sur le maximum de vraisemblance. 
Nous proposons d’appliquer notre approche à la caméra XEMIS2, un prototype préclinique de télescope Compton dédié à l’imagerie TEP 3-photons dans lequel quatre classes de détections partielles viennent s’ajouter aux détections complètes. Sur la base de simulations Monte-Carlo, nous présentons les premières composantes du modèle développé pour la reconstruction multiclasse. 
}
\abstract{
This contribution addresses the problem of image reconstruction of radioactivity distribution for which the available information arises from several classes of data, each associated with a specific combination of detections. 
We introduce a theoretical framework to measure the amount of information brought by each class and we develop an iterative algorithm dedicated to multi-class reconstruction based on maximum likelihood.
We apply our approach to the XEMIS2 camera, a preclinical prototype of a Compton telescope dedicated to 3-photon PET imaging for which four distinct classes of partial detections coexist with the full detection class.
Based on Monte Carlo simulations, we present the first elements of our model.
}
\maketitle
\section{Introduction}
\subsection{Tomographie par émission de positon}
L’imagerie nucléaire permet la visualisation d'un processus biologique par la détection externe des rayonnements émis par les radionucléides préalablement injectés au patient.
Une des méthodes courantes d'imagerie nucléaire est la tomographie par émission de positon (TEP) où les rayonnements détectés sont issus des annihilations de positons émis par le radionucléide avec des électrons naturellement présents dans les tissus.
Le résultat d’une annihilation est l’émission dos-à-dos de deux photons d’énergie initiale égale à 511\,keV.
Ces deux photons sont détectés en coïncidence par une caméra TEP constituée, de façon conventionnelle, par un assemblage de cristaux scintillateurs où chaque couple de cristaux est représenté par une ligne virtuelle appelée ligne de réponse (LOR). Chaque coïncidence détectée permet de localiser la position de l'annihilation quelque part le long de la LOR associée. 
A partir d'une collection de coïncidences, l'étape de reconstruction tomographique vise à produire une image de la distribution radioactive.

Pour améliorer la localisation le long des LORs, la TEP à temps de vol (TOF) mesure la différence des temps de détection des deux photons. 
En utilisant la vitesse de propagation des photons, la différence temporelle permet de calculer la différence de distance parcourue par rapport au centre de la LOR et d'estimer la position de l'annihilation.
La précision de cette méthode dépend de la résolution temporelle du détecteur.

\subsection{Imagerie TEP 3-photons}
Dans cet article, nous nous intéressons à un nouveau mode d'imagerie TEP, dit 3-photons ($3\gamma$), dont l'objectif est d'améliorer la précision de localisation des sources d'émission en utilisant des radionucléides choisis pour émettre quasi simultanément un positon et un photon supplémentaire, appelé troisième photon; ce photon additionnel est ensuite combiné avec les deux photons résultant de l'annihilation du positon.
Le point d’émission du troisième photon est déterminé à l'aide d'une caméra Compton et est localisée à la surface d’un cône de réponse (COR).  
La détection d’un évènement $3\gamma$ permet donc de déterminer ce point à l'intersection d’une LOR et d’un COR.

Plusieurs exemples de caméras précliniques à base de TEP $3\gamma$ existent. 
Le prototype WGI~\cite{yoshida2020} est une caméra TEP modifiée pour réaliser l'imagerie Compton et est en mesure d'opérer des reconstructions séparées à partir de LORs ou de CORs. 
Dans le cas de la détection $3\gamma$, l’origine de l’émission est directement définie comme une des deux intersection LOR/COR et aucune reconstruction d'image n'est appliquée~\cite{mohammadi2022}.
Un autre exemple est la caméra XEMIS2~\cite{gallegomanzano2018}, un télescope Compton dédié à l'imagerie $3\gamma$ à faible activité qui utilise du xénon liquide comme milieu de détection unique et continu. 
Une méthode a été proposée pour convertir les données $3\gamma$ en données TOF afin d'utiliser les méthodes de reconstruction TEP conventionnelles~\cite{giovagnoli2021}. 
Cependant, cette méthode ignore les détections partielles où seulement un ou deux photons sont détectés. 
Même si ces types de détections (nommés classes dans la suite) sont moins informatifs, la combinaison de toutes les classes d'évènements, partielles et complètes, pourrait néanmoins améliorer la qualité de l’image d'activité obtenue en étant intégrées dans un processus de reconstruction multiclasse.

Une méthode de reconstruction multiclasse~\cite{tashima2020b} a été développée pour la caméra WGI. 
L'approche consiste à appliquer un algorithme de reconstruction sur chaque classe d'évènements de manière indépendante, puis de combiner 
les images reconstruites par moyenne pondérée selon l'importance donnée à chaque classe.
Une alternative consiste à effectuer une reconstruction conjointe à partir d'un unique algorithme de reconstruction itératif prenant en compte toutes les classes simultanément. Cette approche a été utilisée pour la caméra Compton MACACO~\cite{munoz2017}, pour laquelle un algorithme prenant en compte les contributions de plusieurs classes d'évènements a été proposé sans justification ni dérivation théorique~\cite{roser2022}.

Cet article présente un cadre général théorique pour la reconstruction conjointe d'une carte d'activité unique à partir d'évènements multiclasse, et nous dérivons un algorithme itératif de reconstruction conjointe basé sur le maximum de vraisemblance qui généralise celui proposé dans~\cite{roser2022}.
Finalement, nous considérons spécifiquement l'imagerie $3\gamma$ avec XEMIS2 comme application. 

\section{Inférence dans le cas multiclasse}
Nous introduisons ici des outils d'inférence dans le cas de données multiclasse, d'une part pour déterminer l'importance relative de chaque classe d'évènement, et d'autre part pour établir un processus de reconstruction itératif selon le principe MLEM~\cite{vardi1985} classique en imagerie TEP. Nous considérons l'image discrète de la distribution de radioactivité $\boldsymbol{\lambda} := \{\lambda_{j} \}_{j\in\llbracket 1,\mathnormal{J} \rrbracket}$ que l'on souhaite reconstruire où chaque $\lambda_{j}$ représente le nombre moyen d'émissions du $j$-ème voxel.  

\subsection{Vraisemblance multiclasse}
Le processus de reconstruction multiclasse est réalisé à partir d'une collection  $\boldsymbol{y} := \left\{\boldsymbol{y}_{n} \right\}_{n\in\llbracket 1,N\rrbracket}$ d'évènements indépendants et identiquement distribués. Chaque évènement $\boldsymbol{y}_{n}$ est un ensemble de mesures contenant les coordonnées des points d'interactions ainsi que les mesures des énergies déposées. Nous notons $K\in\mathbb{N}^{\ast}$ le nombre de classes de détection 
considérées pour la reconstruction.  
\begin{hypo}
    \label{hypo:separability}
    Un évènement $\boldsymbol{y}_{n}$ peut être attribué à une unique classe $k\in\llbracket 1, K\rrbracket$  caractérisée par sa propre densité de probabilité associée à l'occurrence d'un évènement à partir de la distribution $\boldsymbol{\lambda}$.  
\end{hypo}
À partir de l'hypothèse précédente,  l'ensemble des évènements collectés s'exprime comme l'union de $K$ classes:
\begin{align}
    \boldsymbol{y}= \bigcup_{k=1}^{K} \big\{\boldsymbol{y}^{k}_{n} \big\}_{n\in\llbracket 1,\mathnormal{N}^{k} \rrbracket}, 
    \ms \boldsymbol{y}^{k}_{n} \sim p^{k}\big(\boldsymbol{y}^{k}|\boldsymbol{\lambda}\big) ,
    \label{eq:separability}
\end{align}
où $\boldsymbol{y}_{n}^{k} \in \mathbb{R}^{d^{k}}$ est un vecteur de $d^{k}$ mesures continues associées au $n$-ième évènement de classe $k$ détecté et $\sum_{k}N^{k} = N$. 
Dès lors, nous pouvons formuler la proposition suivante: 
\begin{prop}{}
    \label{prop:likelihood}
    La fonction de log-vraisemblance $l(\boldsymbol{\lambda}|\boldsymbol{y})$ s'exprime comme la somme des log-vraisemblances sur chaque classe $l^{k}\big(\boldsymbol{\lambda}|\boldsymbol{y}^{k}\big)$:
    \begin{equation}
    l(\boldsymbol{\lambda}|\boldsymbol{y}) := \sum_{k = 1}^{K}\sum_{n=1}^{N^{k}} \ln p^{k}\big(\boldsymbol{y}^{k}_{n}|\boldsymbol{\lambda}\big) = \sum_{k = 1}^{K} l^{k}\big(\boldsymbol{\lambda}|\boldsymbol{y}^{k}\big).
    \label{eq:loglikelihood}
    \end{equation}
\end{prop}
Cette proposition se démontre en exprimant l'Hypothèse \ref{hypo:separability} à l'aide d'une fonction indicatrice traduisant l'appartenance d'un évènement à une classe.  

En supposant que le processus d'acquisition des évènements de la classe $k$ suit une loi de Poisson de moyenne~$\mu^k=\sum_{j}\lambda_{j}s_{j}^k$, la log-vraisemblance sur chaque classe $k$ s'écrit:
\begin{align*}
    l^{k}\big(\boldsymbol{\lambda}|\boldsymbol{y}^{k}\big) = \sum_{n=1}^{N^{k}} \ln\Bigg( \sum_{j=1}^{J} A_{j}^{k}\big(\boldsymbol{y}^{k}_{n}\big) \lambda_{j}  \Bigg) - N^{k} \ln\Bigg( \sum_{j=1}^{J} \lambda_{j} s_{j}^{k} \Bigg)
\end{align*}
où la \og{}matrice\fg{} système $A^{k}_{j}\left(\boldsymbol{y}^{k}_{n}\right)$ est une fonction continue exprimant la probabilité qu'un évènement de classe $k$ émis depuis le voxel $j$ soit détecté avec les mesures de $\boldsymbol{y}^{k}_{n}$, et où la sensibilité $s_{j}^{k}$, représentant la probabilité qu'un évènement émis depuis le voxel $j$ soit détecté, est définie telle que:
\begin{align}
    s_{j}^{k} := \int_{v\in\Omega^{k}}  A^{k}_{j}(v) \textrm{d}v  \quad \forall j \in \llbracket 1,J\rrbracket,~k \in \llbracket 1,K \rrbracket,
    \label{eq:sensitivity}
\end{align}
où $\Omega^{k}$ est le domaine continu de détection pour la classe $k$.

\subsection{Information de Fisher}
Dans un contexte de données multiclasse, il est intéressant de d'évaluer \emph{a priori} l'intérêt d'intégrer les détections partielles dans la reconstruction en plus des détections complètes, sachant que leur prise en compte représentera un coût de calcul supplémentaire.
L’intérêt de chaque classe peut être mesuré par la quantité d’information qu'elle apporte au sens de la matrice d'information de Fisher (FIM).
En supposant que les densités de probabilité $p^{k}\big(\boldsymbol{y}^{k}|\boldsymbol{\lambda}\big)$ vérifient les bonnes conditions de régularité, l'expression de la FIM proposée dans le cadre de l'imagerie TEP~\cite{parra1998} peut être généralisée pour l'ensemble des classes en utilisant la Proposition~\ref{prop:likelihood} et permet d'exprimer l'information totale comme la somme des informations de toutes les classes, \ie $\mathbf{I}(\boldsymbol{\lambda})=\sum_k
\mathbf{I}^k(\boldsymbol{\lambda})$ avec  
\begin{align}
\label{eq:fisher}
    \big[\mathbf{I}^k(\boldsymbol{\lambda})\big]_{i,j} &= - N^{k}{\Espe}_{\boldsymbol{y}^{k}|\boldsymbol{\lambda}}\left(  \frac{\partial^{2} \ln p^{k}\big(\boldsymbol{y}^{k} | \boldsymbol{\lambda} \big)}{\partial \lambda_{i} \partial\lambda_{j}} \right),\\
        \dfrac{\partial^{2} \ln p^{k}\big(\boldsymbol{y}^{k} | \boldsymbol{\lambda} \big) }{\partial {\lambda_{i}} {\lambda_{j}}} &= \dfrac{s_{i}^{k}s_{j}^{k}}{ \left(\sum_{j'}  \lambda_{j'} s_{j'}^{k}\right)^{2}} -  \dfrac{ A^{k}_{i}\big(\boldsymbol{y}^{k} \big) A^{k}_{j}\big(\boldsymbol{y}^{k} \big)  }{\left(\sum_{j'} A^{k}_{j'}\big(\boldsymbol{y}^{k} \big)  \lambda_{j'}\right)^{2}}.
         \notag
\end{align}
L'indépendance statistique des classes permet de quantifier séparément la contribution de chaque classe $\mathbf{I}^k(\boldsymbol{\lambda}) $ à l'estimation de la distribution; les classes de détections partielles qui présenteront une forte quantité d'information seront alors considérées dans le processus de reconstruction. 

\subsection{Algorithme de reconstruction multiclasse} 
Nous présentons ici une version mode-liste (LM) de l'algorithme MLEM, qui permet de prendre en compte l'aspect continu du milieu de détection pour les modalités d'imagerie TEP~\cite{parra1998} et Compton~\cite{wilderman1998}. 
L'extension du LM-MLEM au cas multiclasse est possible en utilisant la Proposition \ref{prop:likelihood} dans le modèle général de l'algorithme EM~\cite{dempster1977} et permet de décomposer la fonction auxiliaire $Q(\boldsymbol{\lambda}|\boldsymbol{\lambda}^{0})$ en une somme d'approximations inférieures de la vraisemblance pour chaque classe $Q^{k}(\boldsymbol{\lambda}|\boldsymbol{\lambda}^{0})$ vérifiant 
\begin{equation}
    \begin{aligned}
        l(\boldsymbol{\lambda}|\boldsymbol{y}) & \geq Q\big(\boldsymbol{\lambda}|\boldsymbol{\lambda}^{0}\big) =  \sum_{k = 1}^{K} Q^{k}\big(\boldsymbol{\lambda}|\boldsymbol{\lambda}^{0}\big),  \\
        Q\big(\boldsymbol{\lambda}^{0}|\boldsymbol{\lambda}^{0}\big)  & =  l(\boldsymbol{\lambda}^{0}|\boldsymbol{y}) 
    \end{aligned}
    \label{eq:auxi}
\end{equation}
où $\boldsymbol{\lambda}^{0}$ est la distribution radioactive initiale non-négative et non-nulle; la fonction auxiliaire de la classe $k$ s'exprime comme $ Q^{k}\big(\boldsymbol{\lambda}|\boldsymbol{\lambda}^{0}\big) = \Espe_{\boldsymbol{z}^{k} \mid\boldsymbol{y}^{k},\boldsymbol{\lambda}^{0}}\left(\ln p^{k}(\boldsymbol{y}^{k},\boldsymbol{z}^{k}\mid\boldsymbol{\lambda})\right)$ où $\boldsymbol{z}^{k}$ désigne le vecteur de données latentes. 

En appliquant le m\^eme principe que~\cite{parra1998} pour l'expression du LM-MLEM, nous obtenons l'équation de mise à jour de la vraisemblance suivante\footnote{
\textcolor{ao}{
L'expression \eqref{eq:mclmmlem} donnée ici corrige celle qui figure dans les actes du GRETSI'23.
}}: 
\begin{equation}
     \hat{\lambda}^{(t+1)}_{j} := \dfrac{\hat{\lambda}^{(t)}_{j}}{\sum_{k}s^{k}_{j}} \sum_{k = 1}^{K} \sum_{n=1}^{N^{k}} 
     \dfrac{ A^{k}_{j}\left( \boldsymbol{y}_{n}^{k} \right)}{\sum_{j'}
      A^{k}_{j'}\left( \boldsymbol{y}_{n}^{k} \right)
     \hat{\lambda}^{(t)}_{j'}
     + \varepsilon_{n}^{k}
     }
     ,
    \label{eq:mclmmlem}
\end{equation}
où $\varepsilon_{n}^{k}$ modélise les possibles détections diffusées et fortuites sur chaque classe.
L'indépendance statistique des classes permet d’obtenir une expression dans laquelle le facteur multiplicatif s'exprime comme la somme des termes de mise à jour sur chaque classe.

\section{Application à la caméra XEMIS2} 
Les résultats théoriques présentés dans la section précédente sont en cours d'application à la caméra XEMIS2. 
L’émetteur $3\gamma$ actuellement utilisé est le scandium-44  qui émet le troisième photon de manière isotrope avec une énergie initiale de 1157~keV. 
\begin{figure*}[ht]
    \setkeys{Gin}{width=0.85\linewidth}
    \begin{subfigure}{0.20\linewidth}\centering
        \includegraphics{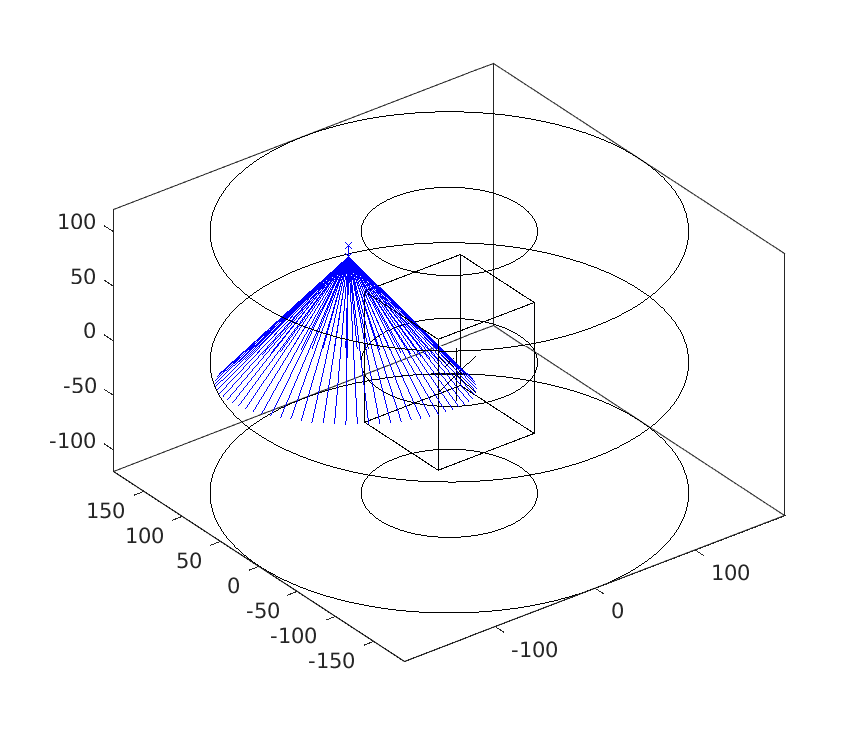}
        \caption{$\CZU$}
    \end{subfigure}%
    \begin{subfigure}{0.20\linewidth}\centering
        \includegraphics{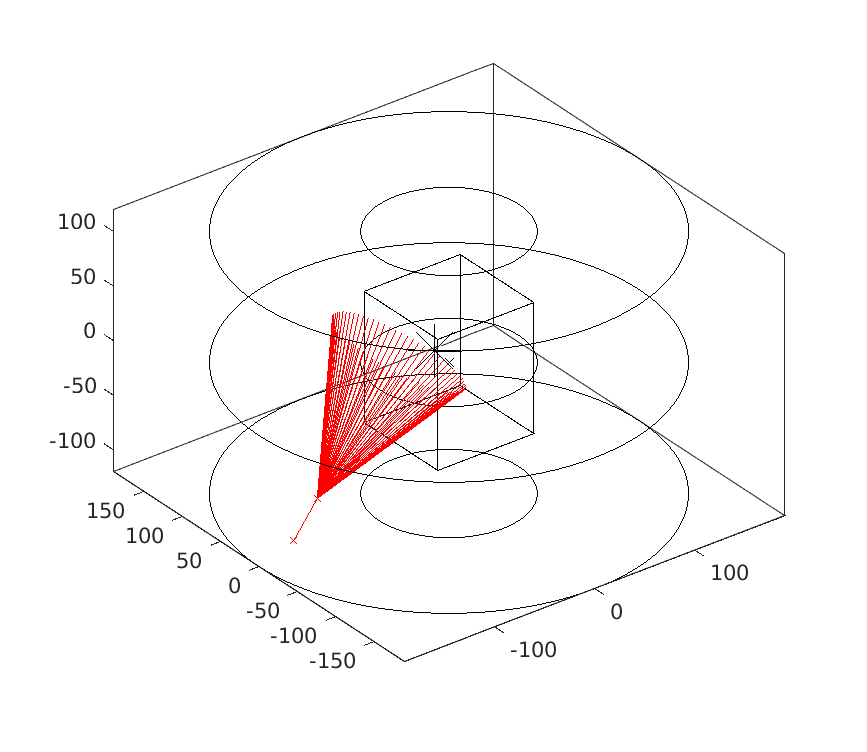}
        \caption{$\CUZ$}
    \end{subfigure}%
    \begin{subfigure}{0.20\linewidth}\centering
        \includegraphics{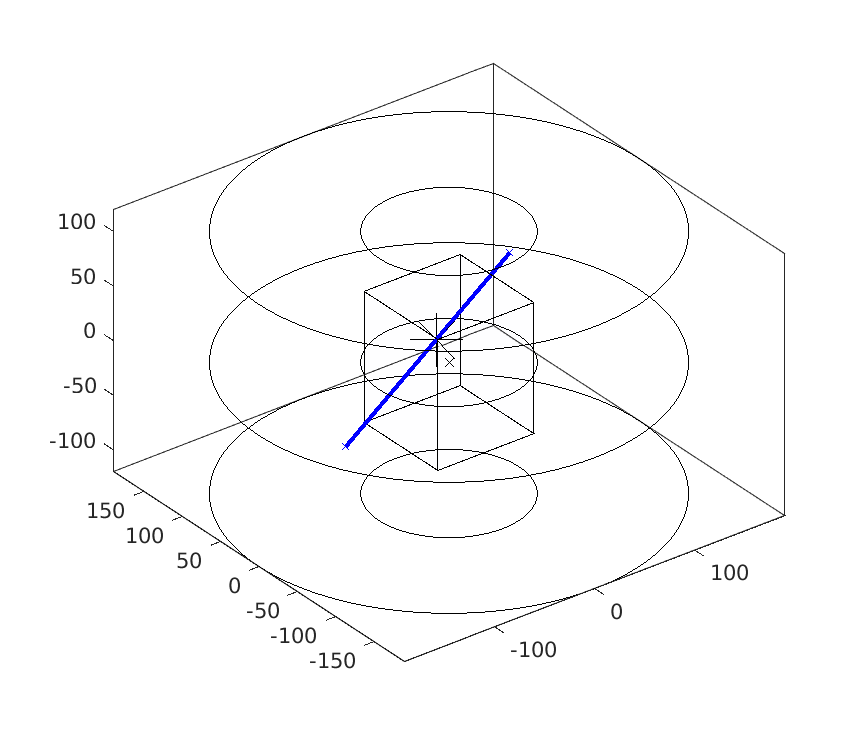}
    \caption{$\CZD$}
    \end{subfigure}%
    \begin{subfigure}{0.20\linewidth}\centering
        \includegraphics{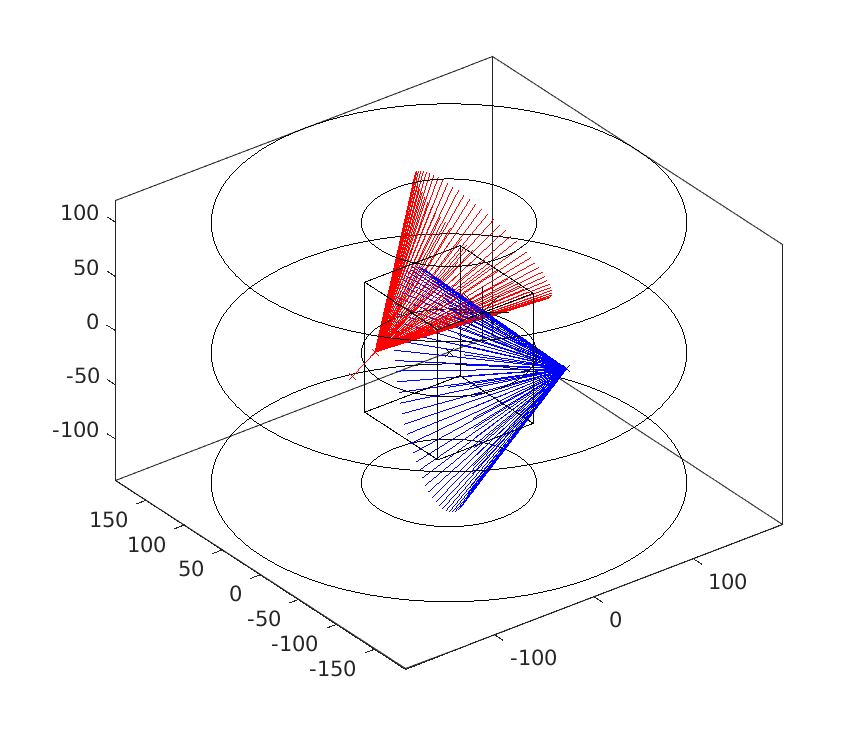}
        \caption{$\CUU$}
    \end{subfigure}%
    \begin{subfigure}{0.20\linewidth}\centering
        \includegraphics{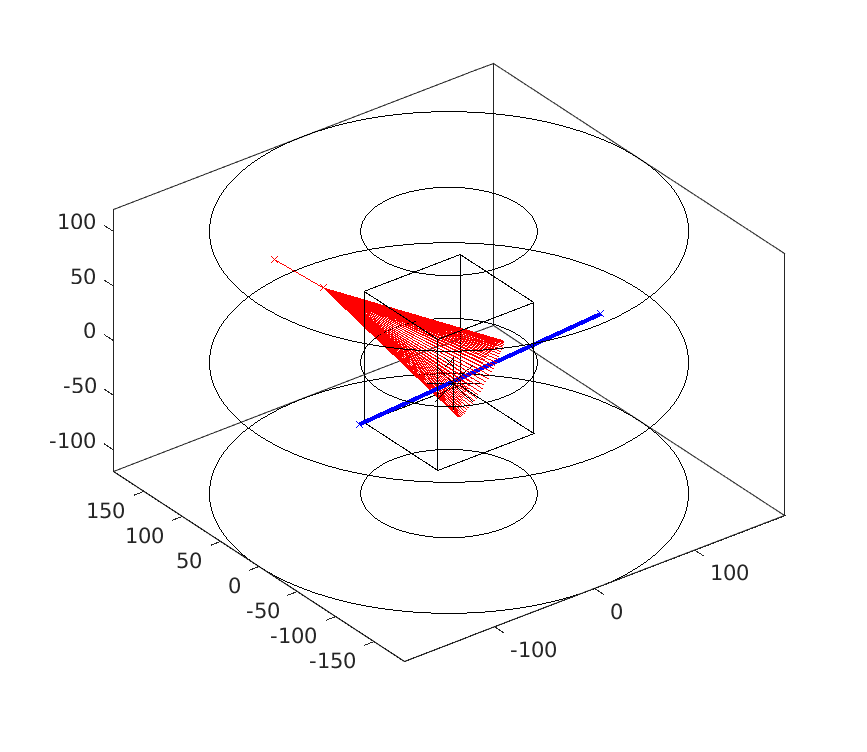}
        \caption{$\CUD$}
    \end{subfigure}
    \caption{Exemples de détections obtenues avec le simulateur Monte Carlo dédié à la caméra XEMIS2. Le milieu de détection continu au xénon liquide est représenté par le cylindre creux et l'image étudiée est représentée par la boîte au centre. Les LORs et CORs bleus sont obtenus à partir de photon(s) d'annihilation et les CORs rouges à partir du troisième photon.} 
    \label{fig:exemples_classes}
\end{figure*}

\subsection{Classes de détections mesurables}
En fonction des énergies des photons détectés dans la caméra XEMIS2, il est possible de définir $K = 5$ classes de détections: 
\begin{itemize}
    \item {$\CZU$} 
    est la classe correspondant à la détection d'un seul photon issu de l'annihilation du positon. La source de l'annihilation est localisée sur un COR ayant un angle de demi-ouverture  $\beta \in [0,\pi]$ tel que: 
    \begin{equation}
        \cos\beta := 
        1- \dfrac{m_{e}c^{2}E_{1}}{E_{0}\left( E_{0}-E_{1}\right)}
        , \quad m_{e}c^{2} = 511 \text{keV}
        \label{eq:comptonangle}
    \end{equation}
    où $E_{0}=511$\,keV  est l'énergie du photon incident et $E_{1}$ est l'énergie déposée dans le milieu de détection durant la diffusion Compton.
    \item {$\CUZ$} 
    correspond à la détection du troisième photon. A l'instar de la classe précédente, le point d'émission appartient à un COR d'angle $\beta$ donné par \eqref{eq:comptonangle} avec $E_{0}=1157$\,keV.
    \item {$\CZD$}
     caractérise les détections de deux photons d'annihilation en coïncidence. Comme pour l'imagerie TEP conventionnelle, la source d'annihilation est localisée sur une LOR joignant les points d'interaction des photons dans le xénon.
    \item {$\CUU$}
    caractérise les détections simultanées d'évènements des classes $\CZU$ et $\CUZ$.
    L'origine de la désintégration est localisée à l'intersection des deux CORs.
    \item {$\CUD$} décrit la classe des détections exprimées comme une combinaison indépendante des classes $\CZD$ et $\CUZ$. L'origine de la désintégration est un point d'intersection LOR/COR. 
\end{itemize}  
Des exemples de détection pour chaque classe d'évènements sont présentés Figure~\ref{fig:exemples_classes}. 

Comme indiqué précédemment, la méthode de reconstruction actuellement proposée pour les images avec XEMIS2~\cite{giovagnoli2021} ne tient compte que de la classe de détection $\CUD$. 
Les quatre autres classes de détections partielles peuvent être intégrées dans le processus de reconstruction avec l’algorithme multiclasse \eqref{eq:mclmmlem}.  

\subsection{Estimation de la sensibilité par classe}
\begin{figure*}[ht]
    \setkeys{Gin}{width=0.75\linewidth}
    \begin{subfigure}{0.185\linewidth}\centering
        \includegraphics{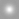}
        \caption{$\CZU$}
    \end{subfigure}%
    \begin{subfigure}{0.185\linewidth}\centering
        \includegraphics{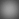}
        \caption{$\CUZ$}
    \end{subfigure}%
    \begin{subfigure}{0.185\linewidth}\centering
        \includegraphics{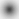}
    \caption{$\CZD$}
    \end{subfigure}%
    \begin{subfigure}{0.185\linewidth}\centering
        \includegraphics{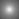}
        \caption{$\CUU$}
    \end{subfigure}%
    \begin{subfigure}{0.185\linewidth}\centering
        \includegraphics{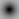}
        \caption{$\CUD$}
    \end{subfigure}
    \begin{subfigure}{0.07\linewidth}\centering
        \includegraphics{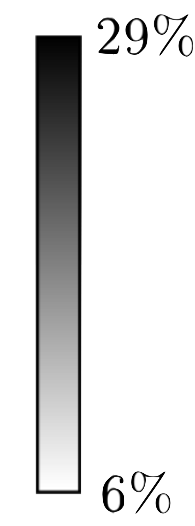}
        \caption*{}
    \end{subfigure}
    \caption{Coupes transaxiales des cartes de sensibilité ($\%$) pour chaque classe d'évènement à la position axiale $z=0$.} 
    \label{fig:axialcrosssection}
\end{figure*}

Le calcul du terme de sensibilité $s^{k}_{j}$, nécessaire pour réaliser une reconstruction, implique l'évaluation d'intégrales multiples et ne peut être dérivé analytiquement (\cf\,\eqref{eq:sensitivity}). 
Pour palier ce problème, nous proposons d'utiliser des approximations des cartes de sensibilité de chaque voxel pour les différentes classes obtenues par simulation Monte Carlo. Un simulateur a été développé permettant de générer des émissions $3\gamma$ et leurs détections dans la caméra XEMIS2 (\href{https://gitlab.com/mlatif/tep3g-pollux}{mlatif/tep3g}). 

Le champ de vue de la caméra est discrétisé en $19 \times 19 \times 24$ voxels de taille $5\times 5 \times 10$\,mm$^{3}$; $M =2\times 10^5$ désintégrations $3\gamma$ ont été générées uniformément dans chaque voxel et la classe d'évènement a été enregistrée pour chaque émission.

\begin{figure}[ht]
  \centering
 \includegraphics[width=.95\columnwidth]{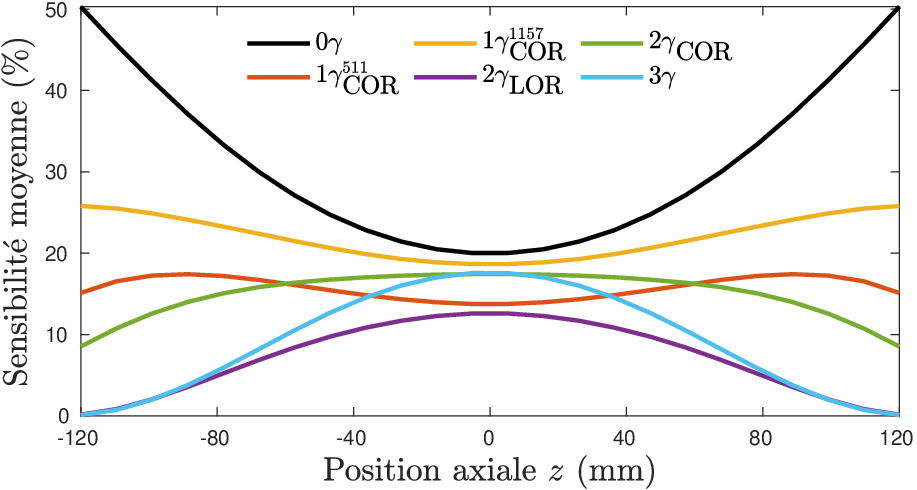}
  \caption{Sensibilité moyenne ($\%$) le long de la direction axiale pour chaque classe d'évènement. La courbe $0\gamma$ correspond à la distribution moyenne des émissions non détectées.} 
  \label{fig:Zprofils}
\end{figure}

La Figure~\ref{fig:axialcrosssection} montre des vues transaxiales des cartes de sensibilité obtenues au centre du champ de vue.
La Figure~\ref{fig:Zprofils} montre des profils des cartes de sensibilité selon la direction axiale.  
Les différentes classes ont des distributions de sensibilité spatialement hétérogènes, avec une certaine complémentarité entre classes. Par exemple, les cartes de sensibilité des classes $\CUD$ et $\CZD$ sont maximales au centre du champ de vue, alors que la tendance est inversée pour la sensibilité des autres classes. 

Le Tableau \ref{tab:sensi_classif} donne la répartition en classe des évènements détectés qui constituent 69\% des émissions. Seulement 13\% des évènements ont conduit à la détection de $\CUD$, ce qui est relativement faible par rapport aux détections partielles. 

Ces observations renforcent l'intérêt de prendre en compte les classes complémentaires de $\CUD$.
\begin{table}[ht]
    \centering
    \small
    \begin{tabular}{lccccc}
    \toprule
    \textbf{Classe} & $\CZU$ & $\CUZ$ & $\CZD$ & $\CUU$ & $\CUD$ \\  
    \midrule
    \textbf{\% de détection} & 22.86 & 32.07 & 10.2 & 21.67 & 13.2 \\
    \bottomrule
    \end{tabular}
    \caption{\label{tab:sensi_classif} Classification des évènements détectés ($\approx$ 69\%) pour des émissions uniformément réparties dans l'ensemble des voxels du champ de vue discrétisé (95 $\times$ 95 $\times$ 240\,mm$^3$). }
\end{table}

\section{Conclusion et perspectives}
Nous avons proposé un cadre théorique pour l'imagerie nucléaire à partir d'évènements multiclasses. En particulier, nous avons spécifié comment évaluer l'information de Fisher apportée par chaque classe, et comment adapter l'algorithme MLEM au cas multiclasse. Cette adaptation peut être étendue à une version pénalisée à la façon de~\cite{depierro1995}, par exemple.

Dans le cadre de l’application à la caméra XEMIS2, les simulations réalisées montrent que les détections partielles sont plus courantes que les détections $3\gamma$ et que certaines de ces classes de détections complètent spatialement la classe $3\gamma$. 
Bien qu'elles soient moins informatives, les détections partielles totalisent 87\% des évènements collectés. Intégrer ces détections dans le processus de reconstruction multiclasse \eqref{eq:mclmmlem} pourrait donc améliorer la qualité de l'estimation de la distribution d'activité des radionucléides.

La prochaine étape de notre projet consistera en l’évaluation de l’information de Fisher~\eqref{eq:fisher} et la mise en œuvre de l'algorithme MLEM multiclasse~\eqref{eq:mclmmlem} dans le logiciel CASToR (Customizable and Advanced Software for Tomographic Reconstruction)~\cite{merlin2018}. Ceci nécessite au préalable d'assurer la prise en charge des évènements Compton dans CASToR, ce qui n'est pas le cas dans la version actuelle. Nous y travaillons actuellement sur la base de travaux proposés par~\cite{maxim2016}.  

\subsection*{Remerciements}
Merci à Nicolas Beaupère pour les précieuses discussions sur la caméra XEMIS2 et le traitement des évènements. 
Ce travail a été en partie financé par le projet NExT Junior Talent TRAC à travers le \og{}Programme d'Investissement Avenir\fg{}.

\printbibliography
\end{document}